\documentstyle[preprint,aps]{revtex}
\begin{document}
\tighten
\title{
      Triple collisions e$^-$\,p$\,{}^7$Be in solar plasma
}
\author{D. E. Monakhov, V. B. Belyaev}
\address{Joint Institute for Nuclear Research, Dubna, 141980, Russia}
\author{S. A. Sofianos, S. A. Rakityansky}
\address{Physics Department, University of South Africa, P.O. Box 392,
Pretoria 0003, South Africa}
\author{W. Sandhas}
\address{Physikalisches Institut, Universit\"{a}t Bonn, 53115 Bonn,
Germany}
\maketitle
\date{\today}

\begin{abstract}
Several nuclear reactions involving the $^7$Be nucleus, not included into
the standard model of the $pp$--chain, are discussed. A qualitative analysis
of their possible influence on the fate of the $^7$Be  in  solar
plasma and of their role in the interpretation of the solar neutrino
experiments is given. As  an example, the reaction rate of the nonradiative
production of $^8$B  in the triple collision
		p + e$^-$ + $^7$Be $\to\ ^8$B+e$^-$
is estimated in  the framework of the adiabatic approximation.
For the solar interior conditions the triple collision
reaction rate is approximately $10^{-4}$
of that for the binary process
	${\rm p}+{}^7{\rm Be} \to  \ {}^8{\rm B}+\gamma$.
\begin{center}
[PACS numbers: 21.45.+v, 95.30.-k, 97.10.Cv]
\end{center}

\end{abstract}
\section{Introduction}
The so--called standard model of the sun has been developed during
several decades by a collective effort of many researchers. At present
it is an elaborate theory which accounts for many, though not all,
observable characteristics of the sun. One of the exceptions is the
flux of neutrinos which is several times higher than what is actually
seen in experiments (for details on the neutrino problem see, for
example, Ref. \cite{Bahcall}). There are several possible reasons
for this discrepancy which are associated either  with the  nuclear
processes involved or with the neutrino propagation towards the earth.

The solar neutrinos emerge from nuclear reactions which occur in
plasma consisting of light nuclei (mainly protons) and electrons. The
sequence of these reactions begins with the proton--proton collision and is,
therefore, called the pp--chain. In the standard model it is assumed
that this chain consists of the following reactions
$$
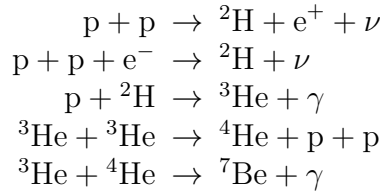
\begin{array}{rcl}
	{\rm p+p} &\to& {}^2{\rm H+e}^+ +\nu \\
	{\rm p+p+e}^- &\to& {}^2{\rm H}+\nu        \\
	{\rm p}+{}^2{\rm H} &\to&  {}^3{\rm He} +\gamma \\
	{}^3{\rm He} +{}^3{\rm He} &\to& {}^4{\rm He + p + p}\\
	{}^3{\rm He}+{}^4{\rm  He} &\to&{}^7{\rm Be}+\gamma \\
\end{array}
$$
and then the $^7$Be is destroyed in two ways
$$
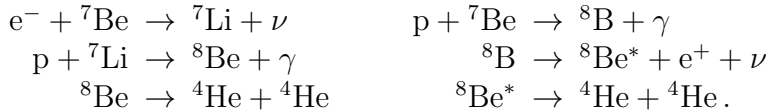
\begin{array}{rclrcl}
	{\rm e^-+{}^7Be} &\to& {}^7{\rm Li} +\nu\qquad \qquad\hfill
	{\rm p}+{}^7{\rm Be} &\to& {}^8{\rm B} +\gamma \hfill \\
	{\rm p}+{}^7{\rm Li} &\to& {}^8{\rm Be}+\gamma
	\qquad \qquad\hfill
	 {}^8{\rm B}&\to& {}^8{\rm Be}{}^* +{\rm e}^+ +\nu\hfill\\
   \qquad\qquad \hfill
       {}^8{\rm Be} &\to& {}^4{\rm He} + {}^4{\rm He}
\qquad \qquad \hfill
       {}^8{\rm Be}^* &\to& {}^4{\rm He} + {}^4{\rm He}\,.
  \qquad \qquad \hfill\\
\end{array}
$$
The fate of  $^7$Be in this chain is of  special interest since a
combined analysis  of all experiments, in which the neutrino flux from the
sun was measured, has come to a paradoxical conclusion: the production of
$^7$Be nuclei (more precisely, the flux of neutrinos due to $^7$Be
reactions) must be strongly suppressed or even be negative \cite{Bah,Deg}.
This implies that something is wrong either in the standard model or in
the experimental data.

In this work we  present  a qualitative as well a quantitative analysis
of the nuclear reactions involving $^7$Be, which may contribute
to a more adequate description of the behavior of this nucleus in the sun.
Our approach differs from the standard model in that we take into
account  also the triple collisions in solar plasma. In contrast the
standard pp--chain  consists of   a sequence of reactions
of two--body initial states, the only exception being the ppe$^-$
collision.

There are two essential differences between the nuclear reactions
caused by  binary and triple collisions. They can be
classified as kinematical and dynamical. The former are related
to the selection rules prevailing in the two-- and three--body
reactions while the latter stem from the inter--dependence of
different nuclear processes. Thus, some binary nuclear reactions
suppressed by  conservation laws (concerning angular momentum,
parity, isospin etc.) can be eased due to the presence of a third
particle. Therefore, the three--body mechanism of such reactions
which is less restricted kinematically, may play a significant
role in the nuclear burning in stellar plasma where the
probability of triple collisions could be quite high due to the
high density of matter. The dynamics of the three particle
motion may also lead to a completely different physical picture:
Processes considered  as independent in binary collisions  become
dependent when  triple collisions are considered. For example,
the ${\rm e}^-+ {}^7{\rm Be}$ and ${\rm p}+{}^7{\rm Be}$ processes,
become dependent to each other when the
${\rm e}^-+{\rm p}+{}^7{\rm Be}$  collision  is taken into account.

The paper is organized as follows. In Sect. II we discuss the role of the
three--body collisions qualitatively. In Sect. III we describe our
formalism and in  IV and V we outline the procedure employed to evaluate
the various  ingredients used to obtain the reaction rate.  In  Sect. VI
we present our results and conclusions. Some details concerning the
derivation of the reaction rate  formula are given in the Appendix.

\section{Binary and triple collisions}
According to the standard model of the sun the $^7$Be nucleus, produced
in the pp--cycle, is destructed  via the following two binary reactions
\begin{eqnarray}
\label{eBeLin}
	{\rm e}^{-}+{}^7{\rm Be} &\to &\ {}^7{\rm Li}+\nu\, ,\\
\label{pBeBg}
	{\rm p}+{}^7{\rm Be} & \to & \ {}^8{\rm B}+\gamma\, .
\end{eqnarray}
Due to the high density of the plasma, proton and electron can always
be found in the vicinity of $^7$Be and form a three--body initial state
\begin{equation}
\label{3init}
	\left|{\rm p+e}^- + {}^7{\rm Be}\right\rangle\, ,
\end{equation}
which can  give rise not only to the above binary transitions
(\ref{eBeLin}) and (\ref{pBeBg})  but also to the processes
\begin{equation}
\label{3body}
	{\rm p+e}^- + {}^7{\rm Be}\longrightarrow\left\{
	    \begin{array}{l}
		^7{\rm Li+p}+\nu\\
		^8{\rm B}+\gamma+{\rm e}^-\\
		^8{\rm B+e}^-\\
\end{array}
\right.\, .
\end{equation}
The first reaction corresponds to the weak transition  (\ref{eBeLin})
provided the nearby proton does not participate in the reaction.
Similarly, the second of these reactions corresponds to the radiative
capture (\ref{pBeBg}) if the electron is a spectator particle. It is
obvious that the three processes (\ref{3body}) are not independent, since
they are generated from the same initial state. Therefore, the
neutrino fluxes related to $^8$B and $^7$Be nuclei are not
independent either. This means that the procedure of comparison of
the Homestake and  Kamiokande experiments, which gave patently
wrong negative amount of $^7$Be, could be incomplete and
should be reconsidered.

The question then arises on how strong is the mutual dependence of
the processes (\ref{3body}). This question can be formulated in an
alternative way: To what extent the two--body subsystems, involved
in (\ref{3body}), can be considered as independent? Formally such
an independence would imply that the total wave function of the
initial state (\ref{3body}) could be written as a direct product of
the wave functions of the e$^-$+$^7$Be, p+${}^7$Be,
and p+e$^-$ subsystems,
\begin{equation}
\label{product}
	\left|{\rm p+e}^- + {}^7{\rm Be}\right\rangle\approx
	\left|{\rm e}^-+^7{\rm Be}\right\rangle\otimes
	\left|{\rm p}+^7{\rm Be}\right\rangle\otimes
	\left|{\rm p+e}^-\right\rangle\, .
\end{equation}
The possibility of such a factorization for three charged particles
was investigated in Refs. \cite{Gar,Brauner}, where it was shown that
the total wave function reduces to a product of the type
(\ref{product}) only when all three particles are far away from each
other. Such a configuration, however, can not contribute much to the
transitions (\ref{3body}), for they are caused by the strong and weak
interactions which vanish at large distances. The description of the
processes (\ref{3body}) as independent needs therefore an additional
substantiation.

There are also other processes which, in principle, can change
the balance between the neutrino fluxes from  $^7$Be and $^8$B
without changing the abundance of $^7$Be nuclei. Two such examples
are
\begin{eqnarray}
\nonumber
	{\rm e}^- +{\rm e}^- + {}^7{\rm Be} &\to&
		\ ^7{\rm Li+ e}^- + \nu\, ,\\
\nonumber
	{\rm p+p} + {}^7{\rm Be} &\to& \ ^8{\rm B+ p}\, .
\end{eqnarray}
In this article, however, we investigate only the processes
leading to the $^8$B formation as described by (\ref{3body}). This
requires the reaction rate of the process
\begin{equation}
\label{nonrad}
	 {\rm p+e}^- +{}{^7{\rm Be}} \to\ {}^8{\rm B + e}^-
\end{equation}
which is discussed next.
%
\section{Reaction rate}
To estimate the reaction rate of (\ref{nonrad}) we assumed that  the
nucleus $^7$Be can be  considered as a paricle and,
therefore, the $^8$B nucleus is a bound state of $^7$Be
with the proton. Of course such an approximation is inadequate if
a full description of the spectrum and  the structure of $^8$B
is required. Since, however, its
ground state is only 0.138 MeV below the (p+$^7$Be)--threshold
\cite{spectrum} (see Fig. 1) this cluster representation should describe it
fairly well.  In the problem under consideration the other
states of $^8$B are not needed since the  collision  occurs at
ultra-low energies of the order of  $\sim$1\,keV (at
temperatures of $\sim 10^7\,{}^{\circ}$K) and thus  transitions via
excited states  can be safely neglected and the formation of $^8$B
proceeds directly from the continuum (p+$^7$Be) through a kind
of Auger transition in which the electron carries away the excess
energy.

The Hamiltonian, which governs the evolution of the three--body initial
state (\ref{3init}), can be written as
\begin{equation}
\label{Ham}
		H=H_0+h_0+V_n+V_e\,,
\end{equation}
where $H_0$ and $h_0$ are the kinetic energy operators associated with the
Jacobi variables $\vec r$ and $\vec\rho$ respectively (see Fig. 2),
$V_n$ is the p$^7$Be potential, which includes strong and Coulomb
parts, and $V_e$ is the  sum of the e$^-$p and e$^-{}^7$Be Coulomb
potentials. The amplitude, describing various transitions in this
three--body system, is proportional to the $T$--operator obeying
the Lippmann--Schwinger equation
\begin{equation}
\label{LSeq}
	T(z)=V_n+V_e +(V_n+V_e)\frac{1}{z-H_0-h_0}T(z)\,.
\end{equation}
The initial asymptotic state (\ref{3init}) is defined by the
p$^7$Be relative momentum  $\vec p$, the total nuclear spin $s$ and its third
component $m_s$, and by the electron momentum with respect to
center of mass of the nuclei $\vec k$. The final asymptotic state
consist of the bound state of a proton and $^7$Be, with spin $\sigma= 2$ and
third component $m_\sigma$, and a freely moving electron with
momentum $\vec k'$ which differs from $\vec k$ not only in its direction but also in its
absolute value, since the electron picked up the energy released in the
nuclear process. We note that we  neglect the spin of the electron
since we assume that its interaction can be satisfactorily described by
purely coulombic potentials without spin--orbit or other forces involving
magnetic moments. Therefore, the reaction (\ref{nonrad}) can be described
as a transition between these asymptotic states,
\begin{equation}
\label{asstran}
	\left|\vec p,s,m_s;\vec k\right\rangle
	\ \stackrel{V_n+V_e}{\Longrightarrow} \
	\left|\Psi^{bound}_{\sigma m_\sigma}, \vec k'\right\rangle\ .
\end{equation}
The corresponding three--body reaction rate ${\cal R}_3$ is given
by (see Ref.\cite{GW})
\begin{equation}
\label{calr}
	{\cal R}_3(\vec p,s,m_s,\vec k\to \sigma m_\sigma,\vec k')
	=\frac{1}{4\pi^2}\delta( E_f-E_i)\left|\left\langle
       \Psi^{bound}_{\sigma m_\sigma},\vec k'\right|T\left|
	\vec p,s,m_s;\vec k\right\rangle
	\right|^2n_{\rm e} n_{\rm p} n_{{}^7{\rm Be}}\, ,
\end{equation}
where the $\delta$--function secures the energy conservation while
$n_{\rm e} n_{\rm p} n_{{}^7{\rm Be}}$ is the product of the
densities of electrons, protons, and ${}^7$Be nuclei in the plasma.
The coefficient $1/4\pi^2$ in Eq. (\ref{calr}) stems from the
following normalizations of the wave functions
\begin{eqnarray*}
	\left\langle\vec p,s,m_s\right.
	\left|\vec p',s',m_s'\right\rangle
	&=&
	  (2\pi)^3 \delta (\vec p-\vec p')
		\delta_{ss'}\delta_{m_s m_s'}\, ,\\
	\left\langle\Psi^{bound}_{\sigma m_\sigma} \right.
	\left|\Psi^{bound}_{\sigma' m_\sigma'} \right\rangle
	&=&
	\delta_{\sigma \sigma'}\delta_{m_\sigma m_\sigma'}\, ,\\
	\left\langle\vec k\right.\left|\vec k'\right\rangle&=&
	 (2\pi )^3\delta (\vec k-\vec k')\ .
\end{eqnarray*}
Both possible values of the p$^7$Be spin, $s=1$ and $s=2$, as well as all
its orientations are represented by different p$^7$Be--pairs in the plasma
with equal probabilities. The momenta $\vec p$ and $\vec k$, however, are
distributed at the temperature $\theta$ according to  Maxwell's law
\begin{eqnarray*}
	N_{\vec p}(\theta) = (2\pi m \kappa_B\theta )^{ - \frac{3}{2}}
		\exp \left(-p^{2}/2m\kappa_B \theta \right)\ ,\\
	N_{\vec k}(\theta) =(2\pi \mu\kappa_B \theta )^{ - \frac{3}{2}}
		\exp \left(-k^{2}/2\mu\kappa_B \theta \right)\ ,
\end{eqnarray*}
where $\kappa_B$ is the Boltzmann constant and the reduced masses $m$ and
$\mu$ are defined by $m^{-1}=m_{\rm p}^{-1}+m_{\rm {}^7Be}^{-1}$ and
$\mu^{-1}=m_{\rm e}^{-1}+(m_{\rm p}+m_{{}^7{\rm Be}})^{-1}$.

We need the average reaction rate  due to all
possible initial states with different $\vec p$, $s$, $m_s$,
$\vec k$ and the total rate of transition
to the final states with any allowed $m_\sigma$ and $\vec k'$. This can be
obtained by averaging the rate (\ref{calr}) over all initial and summing
it over all final quantum numbers. The resulting reaction rate depends only
on the temperature of the plasma
\begin{equation}
\label{bigint1}
	<{\cal R}_3(\theta)> =\sum_{sm_s  m_\sigma}\frac{1}{2s+1}
	\int d^3\vec k'\int d^3\vec p \int d^3\vec k\
	N_{\vec p}(\theta)N_{\vec k}(\theta)
	{\cal R}_3(\vec p,s,m_s,\vec k\to \sigma m_\sigma,\vec k')\ .
\end{equation}
In what follows we consider the various approximations and ingredients
needed to evaluate the reaction rate by means of this general formula.
\section{Adiabatic approximation}
The average kinetic energy of the particles in the plasma,
$<E^{kin}>\sim\kappa_B\theta\approx 1$\,keV,  is the same for nuclei and
electrons, but the velocity of an electron is three orders of magnitude
higher than that of a proton or $^7$Be. Therefore, while the proton
approaches $^7$Be  very slowly, the electron dashes nearby picking
up the energy and leaving the heavy particles in a bound state. Due to
this we can significantly simplify the problem by treating the
relative p$^7$Be motion adiabatically. Such an approximation means
that  the electron bypasses the heavy particles very fast and thus it
has no time to 'observe' any changes in their space position.
Moreover, to participate in such three--body reaction,
the heavy particles must be close to each other when the electron
arrives, because they are too slow as compared to the electron. Hence, while
the electron starts from its asymptotic state $|\vec k\rangle$, the heavy
particles have already interacted to each other via the potential $V_n$
and formed the two--body scattering state
\begin{equation}
\label{scn}
	\left|\vec p,s,m_s\right\rangle
	\ \stackrel{V_n}{\Longrightarrow} \
	\left|\Psi^{scat}_{\vec p s m_s}\right\rangle\ .
\end{equation}
Therefore, instead of the transition (\ref{asstran}) caused by both $V_n$
and $V_e$, in the adiabatic approximation we may consider the transition
\begin{equation}
\label{distort}
	\left|\Psi^{scat}_{\vec psm_s},\vec k\right\rangle
	\ \stackrel{V_e}{\Longrightarrow} \
	\left|\Psi^{bound}_{\sigma m_\sigma},\vec k'\right\rangle\ ,
\end{equation}
where the interaction $V_n$ is taken into account by (\ref{scn}).

Formally, this leads to some simplifications in the Eq. (\ref{LSeq}).
Indeed, since $\Psi^{scat}$ and $\Psi^{bound}$
can be considered as 'excited' and ground states of the nuclear target
respectively, we can describe the transition (\ref{distort}) using
another $T$--matrix defined as
\begin{equation}
\label{LSeq1}
	\tilde T(z)=V_e+V_e\frac{1}{z-h_0-H_A}\tilde T(z)\ ,
\end{equation}
where $H_A=H_0+V_n$ is the total Hamiltonian of the nuclear subsystem.
Within the above approximation we have
\begin{equation}
\label{Tappr}
	\left\langle\Psi^{bound}_{\sigma m_\sigma},\vec k'\right|T\left|
	\vec p,s,m_s;\vec k\right\rangle\approx
	\left\langle\Psi^{bound}_{\sigma m_\sigma},\vec k'\right|
	\tilde T\left|
	\Psi^{scat}_{\vec p, s, m_s},\vec k\right\rangle\ .
\end{equation}
It is known  (see Ref. \cite{FW}) that the adiabatic approximation
is equivalent to the closure approximation that is when all excited
states of the target are assumed to be degenerated and the
Hamiltonian $H_A$ is replaced  by a constant, viz.
\begin{eqnarray*}
	H_A&=&\sum_n {\cal E}_n |\psi_n><\psi_n|
	+ \int dE\,E|\psi_E><\psi_E|\\
	&\approx&{\cal E}_0\left\{
	\sum_n |\psi_n><\psi_n| + \int dE\,|\psi_E>
	<\psi_E|\right\}={\cal E}_0\ .
\end{eqnarray*}
Therefore, in the adiabatic approximation $\tilde T\approx\tilde T^0$,
where $\tilde T^0$ is defined by the following equation
\begin{equation}
\label{LSeq0}
	\tilde T^0(z)=V_e+V_e\frac{1}{z-h_0-{\cal E}_0}\tilde T^0(z)\ .
\end{equation}
In our case ${\cal E}_0$ is the ground state energy of the
$^8$B nucleus, ${\cal E}_0=-0.138$\,MeV. Substituting into
Eq.~(\ref{LSeq0}) the total energy $z=E^{kin}_e+{\cal E}_0$,
where $E^{kin}_e$ is the kinetic energy of the
electron with respect to the center of mass of the heavy particles,
we obtain the Lippmann--Schwinger equation describing the scattering of an
electron from fixed centres. For the solar plasma electrons
$e^2/hv < 1$, which is a sufficient condition to treat Coulomb
interactions in Eq.~(\ref{LSeq0}) perturbatively, i.e.,
$$
	\tilde T^0(z)=V_e+V_e\frac{1}{E^{kin}_e-h_0}V_e+
	V_e\frac{1}{E^{kin}_e-h_0}V_e\frac{1}{E^{kin}_e-h_0}V_e
	+\cdots\, ,
$$
Furthermore, the average potential energy of the electron is of atomic
order of magnitude, $<V_e>\sim 10\ {\rm eV}$, while its kinetic energy
in solar plasma is two orders of magnitude higher,
\mbox{$<E^{kin}_e>\sim 1\ {\rm keV}$}, which implies that the above
iterations should converge very fast. Therefore, we may retain
only the first (Born) term \cite{Taylor}.

Finally, the exact formula (\ref{calr}) for the reaction rate can be
replaced by the following approximate one
\begin{equation}
\label{calra}
	{\cal R}_3(\vec p,s,m_s,\vec k\to \sigma m_\sigma,\vec k')
	=\frac{1}{4\pi^2} \delta( E_f-E_i)\left|\left\langle
	\Psi^{bound}_{\sigma m_\sigma},\vec k'\right|V_e\left|
	\Psi^{scat}_{\vec p s m_s},\vec k\right\rangle
	\right|^2n_{\rm e} n_{\rm p} n_{{}^7{\rm Be}}\ ,
\end{equation}
which can be considered as a three-body generalization of
the distorted wave Born approximation (DWBA) which is widely used
in the theory of the two--body nuclear reactions. The explicit
formula for the reaction rate, which involves
$\Psi^{scat}_{\vec psm_s}(\vec r)$ and $\Psi^{bound}_{\sigma
m_\sigma}(\vec r)$  described in the next section,
is derived in the Appendix.
%
%
\section{Nuclear wave functions}
\subsection{Scattering state}
As mention earlier at the ultra-low energies considered ($\sim$1\,keV)
the collisions between protons and $^7$Be nuclei  are dominated
by the $S$--wave state. Therefore, in the partial wave
expansion of the corresponding scattering wave function we may retain
only the term with zero angular momentum
\begin{equation}
	\Psi^{scat}_{\vec psm_s}(\vec r)=\frac{1}{pr}
	      \sum_{\mu _1 \mu _2}<\frac12\mu_1\frac32\mu_2|sm_s>
	\chi_{\mu_1}^{\rm p}\chi_{\mu_2}^{^7{\rm Be}}\,
	 u^{scat}_p(r)\ ,
\end{equation}
where $\chi_{\mu_1}^{\rm p}$ and $\chi_{\mu_2}^{^7{\rm Be}}$
are the spin functions of the proton and  $^7$Be nucleus.
The $S$--wave radial function $u^{scat}_p(r)$ is a regular
solution of the radial equation
\begin{equation}
\label{radeqs}
	\left\{\frac{{\rm d}^2}{{\rm d}r^2}+p^2-2m
	\left[V_C(r)+V_S(r)\right  ]\right \} u^{scat}_p(r)=0
\end{equation}
with the asymptotic  boundary condition
$$
	u^{scat}_p(r)\mathop{\longrightarrow}
	\limits_{r\to\infty}\sin
		\left[pr-\eta \ln{(2pr)}+\delta_0\right]\ ,
$$
where $\eta $ is the Sommerfeld parameter. The potentials $V_C$
and $V_S$ in Eq. (\ref{radeqs}) describe the Coulomb and strong
p--$^7$Be interactions respectively. In order to take into account
the nonzero size of $^7$Be nucleus we assume that it is a uniformly
charged sphere of radius $R= 2.95$\,fm. Thus  the potential $V_C$ is
\begin{equation}
\label{VC}
	V_{C}(r)=\left\{
		\begin{array}{lcl}
	\displaystyle{1/r} &,& \qquad r\geq R \\
	\displaystyle \frac{3}{2R}-\frac{r^2}{2R^3} &,& \qquad r<R\\
		\end{array}
\right.\ .
\end{equation}
The strong potential $V_S$ is assumed to have a Woods--Saxon form
\cite{Tom}
\begin{equation}
\label{VN}
	  V_S(r)= - \frac{V_0}{1+\exp \left[ (r-R)/a_N \right] }
\end{equation}
with  $V_0=34$\,MeV, $a_N=0.52$\,fm, and the same nuclear radius $R$.
\subsection{Bound state}
The ground state of $^7$Be has the quantum numbers $3/2^-$. Therefore
to construct the ground state of the Boron isotope $^8$B with angular
momentum 2 and positive parity, in the two--body p$^7$Be--model,
the angular momentum $l$ of p$^7$Be relative motion must be $l=1$.
The wave function of $^8$B can then be written as
\begin{equation}
\label{B8}
	\Psi^{bound}_{\sigma m_\sigma}(\vec r)=\frac{1}{r}
	\sum_{\mu_1\mu_2 m_s m }
	<\frac12\mu_1\frac32\mu_2|sm_s>
	<sm_s1m|2m_\sigma>
	\chi_{\mu_1}^{\rm p}
	\chi _{\mu_2}^{^7{\rm Be}}\,\,Y_{1m}(\hat{\vec r})\,
	\phi^{bound}(r)\,.
\end{equation}
The radial wave function $\phi^{bound}(r)$ is a square--integrable
solution of the equation
\begin{equation}
\label{B8eq}
     \left\{\frac{{\rm d}^2}{{\rm d}r^2}+2m{\cal E}
		-\frac{2}{r^{2}}-2m\left[V_C(r)+
	V'_S(r)\right]\right\}\phi^{bound}(r)=0\ ,
\end{equation}
where the potential $ V'_S$  has the same form as (\ref{VN}) but the
strength parameter is slightly modified,  $V'_0= 32.62$\,MeV
to reproduce the bound state of  ${}^8$B, ${\cal E}_0=-0.138$\,MeV.
\section{Results and discussions}
The average production of $^8$B nuclei  per unit volume per second via the
triple collisions (\ref{nonrad}) at a temperature $\theta$ can be written
as (in units  cm$^{-3}$\,sec$^{-1}$)
$$
	<{\cal R}_3(\theta)>=<\Sigma(\theta)>n_{\rm e}
	n_{\rm p}n_{{}^7{\rm Be}}
$$
where $<\Sigma(\theta)>$ is analogous to the so--called transport
cross-section $<\sigma_2 v>$ used in the theory of two--body reactions
\cite{Gau}. Both $<\Sigma (\theta)>$ and $<\sigma_2v>$ are
the reaction rates  corresponding to unit densities of particles.
The explicit form of $<\Sigma(\theta)>$ is given in the Appendix.

The average production of $^8$B via the radiative capture (\ref{pBeBg}) is
(in units cm$^{-3}$\,sec$^{-1}$)
$$
	<{\cal R}_2(\theta)>=<\sigma_2v>n_{\rm p}n_{{}^7{\rm Be}}
$$
Therefore, the ratio of the triple to binary rate depends only on the
density of electrons and the temperature
\begin{equation}
\label{rat}
	\frac{<{\cal R}_3(\theta)>}{<{\cal R}_2(\theta)>}=
	\frac{<\Sigma(\theta)>}{<\sigma_2v>}n_{\rm e}\, .
\end{equation}
We calculated this ratio  using $n_{\rm e} = 100 N_A\, \mbox{cm}^{-3}$
where $N_A=6.022\times10^{23}$ is the Avogadro constant. This density
corresponds to solar plasma conditions~\cite{Bah}. The
results of our calculations for several values of $\theta$ are given in
Table 1. For comparison, this table contains also the absolute values
of the reaction rates of the radiative and nonradiative processes
(\ref{pBeBg}) and (\ref{nonrad}) calculated for $n_{\rm p}
=n_{{}^7{\rm Be}}=N_A$. The data for the radiative proton capture
by ${}^7$Be are taken from Ref. \cite{Gau}.
It is seen that the triple collision (\ref{nonrad})
 plays, apparently, a minor role similarly to the nonradiative
capture reaction  ${\rm e^-+p+{}^2H}\rightarrow {^3{\rm He}}+{\rm e^-}$
 \cite{ours}.
However, at the early stages of the universe when $n_e/N_A \gg 100$,
this reaction,  might have been significant.

The temperature dependence of these reaction rates, normalized
to the same value at $\theta=14\times10^6\,^\circ{\rm K}$, are shown in
Fig. 3. It is seen that the curves for the radiative and nonradiative
processes practically coincide. This is not surprising, for in both cases
the temperature dependence is determined mainly by the same exponential
function stemming from the Maxwellian distribution of the p$^7$Be momentum
$p$ (see formula (\ref{Sigmaf}) in the Appendix). Indeed,
a change of the temperature shifts the maximum of the distribution
which changes the most probable energy of p$^7$Be collision. This in
turn affects the values of
$\Psi^{scat}_{\vec psm_s}(\vec r)$ near the point $r=0$. Since the
Coulomb interaction between the proton and $^7$Be is repulsive and
the collision  energy is very low, the scattering wave function at small
distances depends on the temperature very strongly. This dependence
manifests itself via the corresponding behaviour of the reaction rates
because both the radiative and nonradiative transitions occur at short
distances and their rates depend on the overlapping of
	$\Psi^{scat}_{\vec psm_s}(\vec r)$
with  the bound state wave function
	$\Psi^{bound}_{\sigma m_\sigma}(\vec r)$.

Of course, the energy of the electron also depends on the temperature. This
however has not much influence on the reaction rate because the electron is
attracted to the area of the reaction and its wave function  is always large
at small $\rho$. An additional increase or decrease of it makes, therefore,
no much difference in contrast to the repulsive interaction.

This reasoning leads us to yet another conclusion.
A reaction with a three--body initial state  and with all particles
positively charged, must have very strong temperature dependence.
Thus one should  expect quite different $\theta$--dependence for the
nonradiative $^8$B production with two protons in the initial state
$$
	{\rm p+p} + {}^7{\rm Be} \to \ ^8{\rm B+ p}\, .
$$
Since in this case the adiabatic approximations is not expected
to work well, a more rigorous treatment of this reaction, for example,
using the Faddeev formalism,  is required.

\bigskip
\noindent
{\Large \bf Acknowledgements}\\
Financial support from the Russian Foundation for Basic Research
(grant \# RFBR 96 - 02 - 18678) and
the Foundation for Research Development of
South Africa is greatly appreciated. One of us (VBB) expresses his gratitude
to the Physikalisches Institut der Universit\"at Bonn and
the University of South Africa for their kind hospitality.
\newpage
\appendix
\section{Evaluation of the  transition rate}

The potential $V_e$ describing Coulomb interaction of the electron with the
proton and $^7$Be in the coordinate space is
\begin{equation}
	V_e(\vec{\rho},\vec r)=-\frac{4}{|\vec{\rho}+\beta \vec{r}|}\
	-\ \frac{1}{|\vec{\rho}-\alpha \vec{r}|}
\label{vrp}
\end{equation}
where
$$
	\alpha =\frac{m_{{}^7{\rm Be}}}{m_{{}^7{\rm Be}}+m_{\rm p}}\ ,
	\qquad \beta =1-\alpha\,.
$$
The Jacobi vectors $\vec{\rho}$ and $\vec r$ are depicted in Fig. 2.
According to (\ref{calra}), the matrix element between the initial and
final 'distorted' waves provides the amplitude of the transition
(\ref{distort})
\begin{equation}
	M(\vec p,s,m_s,\vec k\to \sigma m_\sigma,\vec k')
	=\left\langle\Psi^{bound}_{\sigma m_\sigma},\vec k'
	\right|V_e\left|
	\Psi^{scat}_{\vec p s m_s},\vec k\right\rangle=
	\int {\rm d}\vec r \Psi^{bound\,*}_{\sigma m_\sigma}
	(\vec r)U_e(\vec q,\vec r)
	\Psi^{scat}_{\vec psm_s}(\vec r)\ ,
\label{ms1}
\end{equation}
with
\begin{equation}
\label{vqr}
	U_e({\vec q,\vec r})=- \frac{4\pi }{q^{2}}\left[4\exp ({-i\beta
	{\vec q\cdot \vec r})}+\exp(i\alpha {\vec q\cdot \vec r})\right],
\end{equation}
where ${\vec q}=\vec{k}-\vec{k}^{\prime }$  is the  momentum transferred
to the electron, and $ U_e({\vec q,\vec r})$ is the Fourier transform of
the two-center potential (\ref{vrp}) over the electron variable
$\vec{\rho}$. Performing the angular integrations in (\ref{ms1}),
we obtain
\begin{equation}
	M(\vec p,s,m_s,\vec k\to \sigma m_\sigma,\vec k')
	=-\frac{i(4\pi )^{2}}{p q^{2}}\sum_{m}
	   <sm_s1m|\sigma m_\sigma>
	Y_{1m}(\hat{\vec q})\, \left[R(\alpha ,q,p)-4R(\beta ,q,p)\right]\ ,
\label{ms2}
\end{equation}
where the radial integrals are given by
\begin{equation}
	R(\gamma ,q,p)=\int_{0}^{\infty }\phi^{bound}(r)
	u^{scat}_{p}(r)j_{1}(\gamma qr)\,{\rm d}r
\end{equation}
with $j_1$ being the Riccati--Bessel function \cite{abram}.
As  can be seen from  Eq. (\ref{ms2}) and the definition of $\vec{q}$ all
partial waves of the electron motion  contribute to the transition amplitude
$M(\vec p,s,m_s,\vec k\to\sigma m_\sigma,\vec k')$.

By averaging over the initial and summing over the final spin orientations
we obtain
\begin{equation}
\label{msav}
	\frac{1}{2s+1}\sum_{m_s m_\sigma}\left|
	M(\vec p,s,m_s,\vec k\to \sigma m_\sigma,\vec k')\right|^2
	=\frac{5}{2s+1}\ \frac{(4\pi)^3}{16 p^2q^4}\left|R(\alpha ,q,p)
	-4R(\beta,q,p)\right|^{2}\ .
\end{equation}
Then the reaction rate (\ref{bigint1}) can be written as
\begin{equation}
\label{RSigma}
	<{\cal R}_3(\theta)>=<\Sigma(\theta)>n_{\rm e} n_{\rm p}
	n_{{}^7{\rm Be}}\ ,
\end{equation}
where
\begin{equation}
\label{Sigma}
			\begin{array}{rcl}
	<\Sigma(\theta)> &=&\displaystyle\sum_{s}\frac{40\pi^4}{2s+1}
	\int d^3\vec k'\int d^3\vec p \int d^3\vec k\
	N_{\vec p}(\theta)N_{\vec k}(\theta)\\
	&&\\
	&\times&\displaystyle
	\delta \left( \frac{p^2}{2m} + \frac{k^2}{2 \mu}
	- \frac{k^{\prime 2}}{2 \mu}  - {\cal E}_0   \right)
	\frac{1}{p^2q^4}\left|R(\alpha ,q,p)-4R(\beta,q,p)\right|^{2}\,.\\
			\end{array}
\end{equation}
The integration over $\vec k^\prime$ in the above formula can be
performed using $\vec q = \vec k^\prime - \vec k $ instead of $\vec k'$
in the integration. Then the $\delta$--function  facilitates
the integration over the solid angle of  $\vec q $.

Finally, the unit density reaction rate $<\Sigma(\theta)>$ is expressed in
the form of a two--dimensional integral
\begin{equation}
\label{Sigmaf}
			\begin{array}{rcl}
	<\Sigma(\theta)> &=&\displaystyle\sum_{s}
	\frac{10(2\pi)^4}{2s+1}\frac{\sqrt{\mu}}{\theta^2}
	\int^\infty_0\ {\rm d}p \exp \left ( -\ \frac {p^2}{2\theta}
	- \frac {2\pi}{p} \right )\\
		&&\\
	&\times&\displaystyle
	\int^\infty_0\ \frac{{\rm d}q}{q^3} \exp \left [ -\
	\frac {\mu}{2\theta q^2} \left ( \frac {p^2}{2}
	- \frac {q^2}{2\mu} - {\cal E}_0\right )^2 \right ]
	\left[Q(\alpha,p,q)-4Q(\beta,p,q)\right]^2\\
			\end{array}
\end{equation}
with
$$
	Q(\alpha,p,q) = \int^\infty_0\ \phi^{bound}(r)
	j_1(\alpha qr) u^{scat}_p (r)
	\exp\left(\frac {\pi}{p}\right) {\rm d}r\ .
$$



\newpage

\begin{table}
\label{tableI}

\begin{tabular}{|c|c|c|c|}
$\phantom{--}\theta_6\phantom{--}$
	& $<\sigma_2v>N_A$                       & $<\Sigma(\theta)>N_A^2$
	& $<{\cal R}_3(\theta)>/<{\cal R}_2(\theta)>$ \\
 $10^6\,^\circ$K.
	&    ( cm$^3$\,mol$^{-1}$\,sec$^{-1}$) \cite{Gau}
	&    ( cm$^3$\,mol$^{-1}$\,sec$^{-1}$)   &   \\
\hline
 14&1.73 $\times 10^{-12}$&1.83 $\times 10^{-16}$ &1.06$\times 10^{-4}$  \\
 15&4.34 $\times 10^{-12}$&4.43 $\times 10^{-16}$ &1.02$\times 10^{-4}$  \\
 16&10.1 $\times 10^{-12}$&9.80 $\times 10^{-16}$ &0.97$\times 10^{-4}$  \\
 18&44.7 $\times 10^{-12}$&42.0 $\times 10^{-16}$ &0.94$\times 10^{-4}$  \\
 20&161.0$\times 10^{-12}$&143.3$\times 10^{-16}$ &0.89$\times 10^{-4}$  \\
\end{tabular}
\vspace{1cm}
\caption{Temperature dependence of the reaction rates
for the radiative and nonradiative formation of $^8$B
in solar plasma.}
\end{table}

\newpage
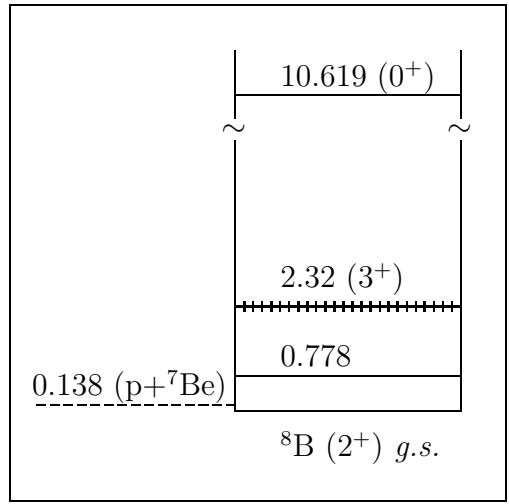
\begin{figure}

\begin{center}
\unitlength=0.6mm
\begin{picture}(110,110)
\put(0,0){\line(1,0){110}}
\put(0,0){\line(0,1){110}}
\put(110,0){\line(0,1){110}}
\put(0,110){\line(1,0){110}}
\put(50,20){%
\begin{picture}(0,0)
\put(0,0){\line(1,0){50}}
\multiput(0,0)(50,0){2}{%
\begin{picture}(0,0)
\put(0,0){\line(0,1){60}}
\put(-3,61){$\sim$}
\put(0,65){\line(0,1){15}}
\end{picture}}
\put(0,7.78){\line(1,0){50}}
\put(0,23.2){\line(1,0){50}}
\put(0,70){\line(1,0){50}}
\put(10,10){0.778}
\put(10,27.5){2.32\ (3$^+$)}
\multiput(0,22.2)(2,0){26}{\line(0,1){2}}
\put(10,72.2){10.619\ (0$^+$)}
\put(10,-10){${}^8$B\ (2$^+$)\ {\it g.s.}}
\multiput(0,1.38)(-3,0){15}{\line(-1,0){2}}
\put(-45,3.6){0.138\ (p+$^7$Be)}
\end{picture}}
\end{picture}
\end{center}
\caption{Spectrum of  $^8$B nucleus. Energy levels are in MeV.}
\end{figure}
\vspace*{2cm}
\begin{figure}

\begin{center}
\unitlength=0.6mm
\begin{picture}(140,80)
\put(0,0){\line(1,0){140}}
\put(0,0){\line(0,1){80}}
\put(140,0){\line(0,1){80}}
\put(0,80){\line(1,0){140}}
\put(25,50){\vector(3,-1){56.5}}
\put(100,58.59){\vector(-2,-3){23}}
\put(25,50){\circle*{2}}
\put(101.2,60.2){\circle{4}}
\put(74,20){\circle{10}}
\put(15,49){e$^-$}
\put(107,60){p}
\put(83,17){$^7$Be}
\put(48,33){$\vec\rho$}
\put(83,42){$\vec r$}
\end{picture}
\end{center}
\caption{The Jacobi vectors.}
\end{figure}
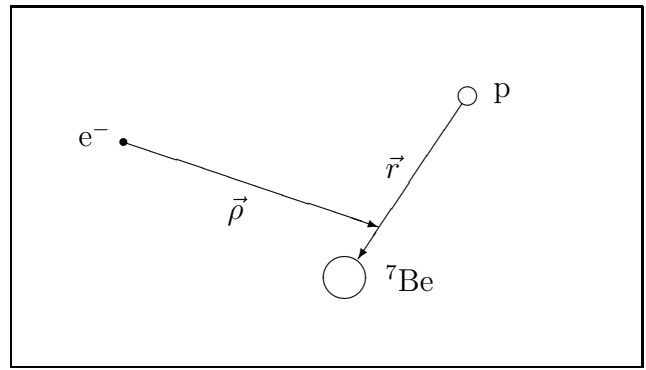
\vspace*{2cm}
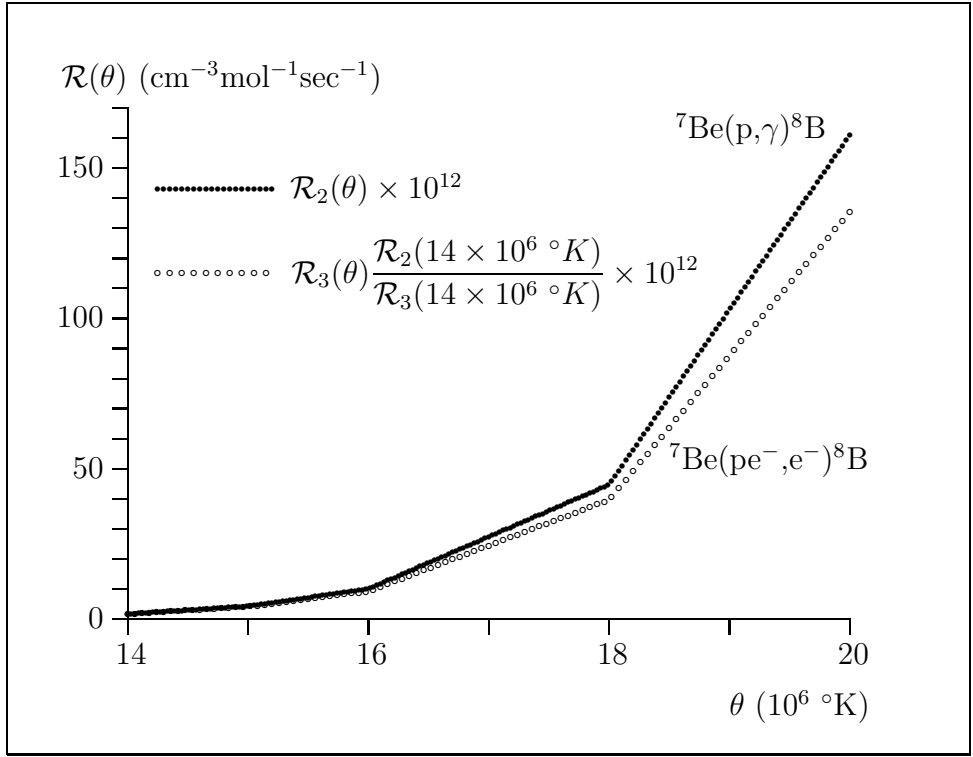
\begin{figure}
\begin{center}
\unitlength=0.4mm
\begin{picture}(320,250)
\put(0,0){\line(1,0){320}}
\put(0,0){\line(0,1){250}}
\put(320,0){\line(0,1){250}}
\put(0,250){\line(1,0){320}}
\put(40,45){%
\begin{picture}(0,0)
\put(0,0){\line(1,0){240}}
\put(0,0){\line(0,1){170}}
\multiput(0,0)(40,0){7}{\line(0,-1){5}}
\multiput(0,0)(0,10){18}{\line(-1,0){5}}
\put(-4,-15){14}
\put(76,-15){16}
\put(156,-15){18}
\put(236,-15){20}
\put(-8,-3){\llap{0}}
\put(-8,47){\llap{50}}
\put(-8,97){\llap{100}}
\put(-8,147){\llap{150}}
\put(200,-32){$\theta$\  ($10^6\ {}^{\circ}$K)}
\put(-22,177){${\cal R}(\theta)$\ (cm$^{-3}$mol$^{-1}$sec$^{-1}$)}
\multiput(10,143)(2,0){20}{\circle*{2}}
\multiput(10,115)(4,0){10}{\circle{2}}
\put(54,140){${\cal R}_2(\theta)\times 10^{12}$}
\put(51,112){
$ {\cal R}_3(\theta)
\displaystyle \frac{    {\cal R}_2(14\times 10^6\ {}^{\circ}K)     }
{    {\cal R}_3(14\times 10^6\ {}^{\circ}K)    }
\times 10^{12}$
}
\put(180,50){${}^7$Be(pe$^-$,e$^-$)${}^8$B}
\put(182,160){$^7$Be(p,$\gamma$)$^8$B}
\put(  .00000    ,  1.7300    ){\circle*{2}}
\put(  1.2000    ,  1.8083    ){\circle*{2}}
\put(  2.4000    ,  1.8866    ){\circle*{2}}
\put(  3.6000    ,  1.9649    ){\circle*{2}}
\put(  4.8000    ,  2.0432    ){\circle*{2}}
\put(  6.0000    ,  2.1215    ){\circle*{2}}
\put(  7.2000    ,  2.1998    ){\circle*{2}}
\put(  8.4000    ,  2.2781    ){\circle*{2}}
\put(  9.6000    ,  2.3564    ){\circle*{2}}
\put(  10.800    ,  2.4347    ){\circle*{2}}
\put(  12.000    ,  2.5130    ){\circle*{2}}
\put(  13.200    ,  2.5913    ){\circle*{2}}
\put(  14.400    ,  2.6696    ){\circle*{2}}
\put(  15.600    ,  2.7479    ){\circle*{2}}
\put(  16.800    ,  2.8262    ){\circle*{2}}
\put(  18.000    ,  2.9045    ){\circle*{2}}
\put(  19.200    ,  2.9828    ){\circle*{2}}
\put(  20.400    ,  3.0611    ){\circle*{2}}
\put(  21.600    ,  3.1394    ){\circle*{2}}
\put(  22.800    ,  3.2177    ){\circle*{2}}
\put(  24.000    ,  3.2960    ){\circle*{2}}
\put(  25.200    ,  3.3743    ){\circle*{2}}
\put(  26.400    ,  3.4526    ){\circle*{2}}
\put(  27.600    ,  3.5309    ){\circle*{2}}
\put(  28.800    ,  3.6092    ){\circle*{2}}
\put(  30.000    ,  3.6875    ){\circle*{2}}
\put(  31.200    ,  3.7658    ){\circle*{2}}
\put(  32.400    ,  3.8441    ){\circle*{2}}
\put(  33.600    ,  3.9224    ){\circle*{2}}
\put(  34.800    ,  4.0007    ){\circle*{2}}
\put(  36.000    ,  4.0790    ){\circle*{2}}
\put(  37.200    ,  4.1573    ){\circle*{2}}
\put(  38.400    ,  4.2356    ){\circle*{2}}
\put(  39.600    ,  4.3139    ){\circle*{2}}
\put(  40.800    ,  4.4552    ){\circle*{2}}
\put(  42.000    ,  4.6280    ){\circle*{2}}
\put(  43.200    ,  4.8008    ){\circle*{2}}
\put(  44.400    ,  4.9736    ){\circle*{2}}
\put(  45.600    ,  5.1464    ){\circle*{2}}
\put(  46.800    ,  5.3192    ){\circle*{2}}
\put(  48.000    ,  5.4920    ){\circle*{2}}
\put(  49.200    ,  5.6648    ){\circle*{2}}
\put(  50.400    ,  5.8376    ){\circle*{2}}
\put(  51.600    ,  6.0104    ){\circle*{2}}
\put(  52.800    ,  6.1832    ){\circle*{2}}
\put(  54.000    ,  6.3560    ){\circle*{2}}
\put(  55.200    ,  6.5288    ){\circle*{2}}
\put(  56.400    ,  6.7016    ){\circle*{2}}
\put(  57.600    ,  6.8744    ){\circle*{2}}
\put(  58.800    ,  7.0472    ){\circle*{2}}
\put(  60.000    ,  7.2200    ){\circle*{2}}
\put(  61.200    ,  7.3928    ){\circle*{2}}
\put(  62.400    ,  7.5656    ){\circle*{2}}
\put(  63.600    ,  7.7384    ){\circle*{2}}
\put(  64.800    ,  7.9112    ){\circle*{2}}
\put(  66.000    ,  8.0840    ){\circle*{2}}
\put(  67.200    ,  8.2568    ){\circle*{2}}
\put(  68.400    ,  8.4296    ){\circle*{2}}
\put(  69.600    ,  8.6024    ){\circle*{2}}
\put(  70.800    ,  8.7752    ){\circle*{2}}
\put(  72.000    ,  8.9480    ){\circle*{2}}
\put(  73.200    ,  9.1208    ){\circle*{2}}
\put(  74.400    ,  9.2936    ){\circle*{2}}
\put(  75.600    ,  9.4664    ){\circle*{2}}
\put(  76.800    ,  9.6392    ){\circle*{2}}
\put(  78.000    ,  9.8120    ){\circle*{2}}
\put(  79.200    ,  9.9848    ){\circle*{2}}
\put(  80.400    ,  10.273    ){\circle*{2}}
\put(  81.600    ,  10.792    ){\circle*{2}}
\put(  82.800    ,  11.311    ){\circle*{2}}
\put(  84.000    ,  11.830    ){\circle*{2}}
\put(  85.200    ,  12.349    ){\circle*{2}}
\put(  86.400    ,  12.868    ){\circle*{2}}
\put(  87.600    ,  13.387    ){\circle*{2}}
\put(  88.800    ,  13.906    ){\circle*{2}}
\put(  90.000    ,  14.425    ){\circle*{2}}
\put(  91.200    ,  14.944    ){\circle*{2}}
\put(  92.400    ,  15.463    ){\circle*{2}}
\put(  93.600    ,  15.982    ){\circle*{2}}
\put(  94.800    ,  16.501    ){\circle*{2}}
\put(  96.000    ,  17.020    ){\circle*{2}}
\put(  97.200    ,  17.539    ){\circle*{2}}
\put(  98.400    ,  18.058    ){\circle*{2}}
\put(  99.600    ,  18.577    ){\circle*{2}}
\put(  100.80    ,  19.096    ){\circle*{2}}
\put(  102.00    ,  19.615    ){\circle*{2}}
\put(  103.20    ,  20.134    ){\circle*{2}}
\put(  104.40    ,  20.653    ){\circle*{2}}
\put(  105.60    ,  21.172    ){\circle*{2}}
\put(  106.80    ,  21.691    ){\circle*{2}}
\put(  108.00    ,  22.210    ){\circle*{2}}
\put(  109.20    ,  22.729    ){\circle*{2}}
\put(  110.40    ,  23.248    ){\circle*{2}}
\put(  111.60    ,  23.767    ){\circle*{2}}
\put(  112.80    ,  24.286    ){\circle*{2}}
\put(  114.00    ,  24.805    ){\circle*{2}}
\put(  115.20    ,  25.324    ){\circle*{2}}
\put(  116.40    ,  25.843    ){\circle*{2}}
\put(  117.60    ,  26.362    ){\circle*{2}}
\put(  118.80    ,  26.881    ){\circle*{2}}
\put(  120.00    ,  27.400    ){\circle*{2}}
\put(  121.20    ,  27.919    ){\circle*{2}}
\put(  122.40    ,  28.438    ){\circle*{2}}
\put(  123.60    ,  28.957    ){\circle*{2}}
\put(  124.80    ,  29.476    ){\circle*{2}}
\put(  126.00    ,  29.995    ){\circle*{2}}
\put(  127.20    ,  30.514    ){\circle*{2}}
\put(  128.40    ,  31.033    ){\circle*{2}}
\put(  129.60    ,  31.552    ){\circle*{2}}
\put(  130.80    ,  32.071    ){\circle*{2}}
\put(  132.00    ,  32.590    ){\circle*{2}}
\put(  133.20    ,  33.109    ){\circle*{2}}
\put(  134.40    ,  33.628    ){\circle*{2}}
\put(  135.60    ,  34.147    ){\circle*{2}}
\put(  136.80    ,  34.666    ){\circle*{2}}
\put(  138.00    ,  35.185    ){\circle*{2}}
\put(  139.20    ,  35.704    ){\circle*{2}}
\put(  140.40    ,  36.223    ){\circle*{2}}
\put(  141.60    ,  36.742    ){\circle*{2}}
\put(  142.80    ,  37.261    ){\circle*{2}}
\put(  144.00    ,  37.780    ){\circle*{2}}
\put(  145.20    ,  38.299    ){\circle*{2}}
\put(  146.40    ,  38.818    ){\circle*{2}}
\put(  147.60    ,  39.337    ){\circle*{2}}
\put(  148.80    ,  39.856    ){\circle*{2}}
\put(  150.00    ,  40.375    ){\circle*{2}}
\put(  151.20    ,  40.894    ){\circle*{2}}
\put(  152.40    ,  41.413    ){\circle*{2}}
\put(  153.60    ,  41.932    ){\circle*{2}}
\put(  154.80    ,  42.451    ){\circle*{2}}
\put(  156.00    ,  42.970    ){\circle*{2}}
\put(  157.20    ,  43.489    ){\circle*{2}}
\put(  158.40    ,  44.008    ){\circle*{2}}
\put(  159.60    ,  44.527    ){\circle*{2}}
\put(  160.80    ,  45.863    ){\circle*{2}}
\put(  162.00    ,  47.607    ){\circle*{2}}
\put(  163.20    ,  49.352    ){\circle*{2}}
\put(  164.40    ,  51.097    ){\circle*{2}}
\put(  165.60    ,  52.841    ){\circle*{2}}
\put(  166.80    ,  54.586    ){\circle*{2}}
\put(  168.00    ,  56.330    ){\circle*{2}}
\put(  169.20    ,  58.074    ){\circle*{2}}
\put(  170.40    ,  59.819    ){\circle*{2}}
\put(  171.60    ,  61.563    ){\circle*{2}}
\put(  172.80    ,  63.308    ){\circle*{2}}
\put(  174.00    ,  65.053    ){\circle*{2}}
\put(  175.20    ,  66.797    ){\circle*{2}}
\put(  176.40    ,  68.541    ){\circle*{2}}
\put(  177.60    ,  70.286    ){\circle*{2}}
\put(  178.80    ,  72.030    ){\circle*{2}}
\put(  180.00    ,  73.775    ){\circle*{2}}
\put(  181.20    ,  75.520    ){\circle*{2}}
\put(  182.40    ,  77.264    ){\circle*{2}}
\put(  183.60    ,  79.009    ){\circle*{2}}
\put(  184.80    ,  80.753    ){\circle*{2}}
\put(  186.00    ,  82.497    ){\circle*{2}}
\put(  187.20    ,  84.242    ){\circle*{2}}
\put(  188.40    ,  85.986    ){\circle*{2}}
\put(  189.60    ,  87.731    ){\circle*{2}}
\put(  190.80    ,  89.476    ){\circle*{2}}
\put(  192.00    ,  91.220    ){\circle*{2}}
\put(  193.20    ,  92.965    ){\circle*{2}}
\put(  194.40    ,  94.709    ){\circle*{2}}
\put(  195.60    ,  96.453    ){\circle*{2}}
\put(  196.80    ,  98.198    ){\circle*{2}}
\put(  198.00    ,  99.943    ){\circle*{2}}
\put(  199.20    ,  101.69    ){\circle*{2}}
\put(  200.40    ,  103.43    ){\circle*{2}}
\put(  201.60    ,  105.18    ){\circle*{2}}
\put(  202.80    ,  106.92    ){\circle*{2}}
\put(  204.00    ,  108.67    ){\circle*{2}}
\put(  205.20    ,  110.41    ){\circle*{2}}
\put(  206.40    ,  112.15    ){\circle*{2}}
\put(  207.60    ,  113.90    ){\circle*{2}}
\put(  208.80    ,  115.64    ){\circle*{2}}
\put(  210.00    ,  117.39    ){\circle*{2}}
\put(  211.20    ,  119.13    ){\circle*{2}}
\put(  212.40    ,  120.88    ){\circle*{2}}
\put(  213.60    ,  122.62    ){\circle*{2}}
\put(  214.80    ,  124.37    ){\circle*{2}}
\put(  216.00    ,  126.11    ){\circle*{2}}
\put(  217.20    ,  127.85    ){\circle*{2}}
\put(  218.40    ,  129.60    ){\circle*{2}}
\put(  219.60    ,  131.34    ){\circle*{2}}
\put(  220.80    ,  133.09    ){\circle*{2}}
\put(  222.00    ,  134.83    ){\circle*{2}}
\put(  223.20    ,  136.58    ){\circle*{2}}
\put(  224.40    ,  138.32    ){\circle*{2}}
\put(  225.60    ,  140.07    ){\circle*{2}}
\put(  226.80    ,  141.81    ){\circle*{2}}
\put(  228.00    ,  143.56    ){\circle*{2}}
\put(  229.20    ,  145.30    ){\circle*{2}}
\put(  230.40    ,  147.04    ){\circle*{2}}
\put(  231.60    ,  148.79    ){\circle*{2}}
\put(  232.80    ,  150.53    ){\circle*{2}}
\put(  234.00    ,  152.28    ){\circle*{2}}
\put(  235.20    ,  154.02    ){\circle*{2}}
\put(  236.40    ,  155.77    ){\circle*{2}}
\put(  237.60    ,  157.51    ){\circle*{2}}
\put(  238.80    ,  159.26    ){\circle*{2}}
\put(  240.00    ,  161.00    ){\circle*{2}}
\put(  .00000    ,  1.7300    ){\circle{2}}
\put(  2.4000    ,  1.8775    ){\circle{2}}
\put(  4.8000    ,  2.0250    ){\circle{2}}
\put(  7.2000    ,  2.1724    ){\circle{2}}
\put(  9.6000    ,  2.3199    ){\circle{2}}
\put(  12.000    ,  2.4674    ){\circle{2}}
\put(  14.400    ,  2.6149    ){\circle{2}}
\put(  16.800    ,  2.7623    ){\circle{2}}
\put(  19.200    ,  2.9098    ){\circle{2}}
\put(  21.600    ,  3.0573    ){\circle{2}}
\put(  24.000    ,  3.2048    ){\circle{2}}
\put(  26.400    ,  3.3522    ){\circle{2}}
\put(  28.800    ,  3.4997    ){\circle{2}}
\put(  31.200    ,  3.6472    ){\circle{2}}
\put(  33.600    ,  3.7947    ){\circle{2}}
\put(  36.000    ,  3.9421    ){\circle{2}}
\put(  38.400    ,  4.0896    ){\circle{2}}
\put(  40.800    ,  4.2895    ){\circle{2}}
\put(  43.200    ,  4.5940    ){\circle{2}}
\put(  45.600    ,  4.8986    ){\circle{2}}
\put(  48.000    ,  5.2032    ){\circle{2}}
\put(  50.400    ,  5.5078    ){\circle{2}}
\put(  52.800    ,  5.8124    ){\circle{2}}
\put(  55.200    ,  6.1170    ){\circle{2}}
\put(  57.600    ,  6.4216    ){\circle{2}}
\put(  60.000    ,  6.7262    ){\circle{2}}
\put(  62.400    ,  7.0308    ){\circle{2}}
\put(  64.800    ,  7.3354    ){\circle{2}}
\put(  67.200    ,  7.6400    ){\circle{2}}
\put(  69.600    ,  7.9446    ){\circle{2}}
\put(  72.000    ,  8.2492    ){\circle{2}}
\put(  74.400    ,  8.5538    ){\circle{2}}
\put(  76.800    ,  8.8584    ){\circle{2}}
\put(  79.200    ,  9.1629    ){\circle{2}}
\put(  81.600    ,  9.8733    ){\circle{2}}
\put(  84.000    ,  10.787    ){\circle{2}}
\put(  86.400    ,  11.700    ){\circle{2}}
\put(  88.800    ,  12.613    ){\circle{2}}
\put(  91.200    ,  13.526    ){\circle{2}}
\put(  93.600    ,  14.439    ){\circle{2}}
\put(  96.000    ,  15.353    ){\circle{2}}
\put(  98.400    ,  16.266    ){\circle{2}}
\put(  100.80    ,  17.179    ){\circle{2}}
\put(  103.20    ,  18.092    ){\circle{2}}
\put(  105.60    ,  19.005    ){\circle{2}}
\put(  108.00    ,  19.919    ){\circle{2}}
\put(  110.40    ,  20.832    ){\circle{2}}
\put(  112.80    ,  21.745    ){\circle{2}}
\put(  115.20    ,  22.658    ){\circle{2}}
\put(  117.60    ,  23.571    ){\circle{2}}
\put(  120.00    ,  24.485    ){\circle{2}}
\put(  122.40    ,  25.398    ){\circle{2}}
\put(  124.80    ,  26.311    ){\circle{2}}
\put(  127.20    ,  27.224    ){\circle{2}}
\put(  129.60    ,  28.138    ){\circle{2}}
\put(  132.00    ,  29.051    ){\circle{2}}
\put(  134.40    ,  29.964    ){\circle{2}}
\put(  136.80    ,  30.877    ){\circle{2}}
\put(  139.20    ,  31.790    ){\circle{2}}
\put(  141.60    ,  32.704    ){\circle{2}}
\put(  144.00    ,  33.617    ){\circle{2}}
\put(  146.40    ,  34.530    ){\circle{2}}
\put(  148.80    ,  35.443    ){\circle{2}}
\put(  151.20    ,  36.356    ){\circle{2}}
\put(  153.60    ,  37.270    ){\circle{2}}
\put(  156.00    ,  38.183    ){\circle{2}}
\put(  158.40    ,  39.096    ){\circle{2}}
\put(  160.80    ,  40.663    ){\circle{2}}
\put(  163.20    ,  43.535    ){\circle{2}}
\put(  165.60    ,  46.408    ){\circle{2}}
\put(  168.00    ,  49.281    ){\circle{2}}
\put(  170.40    ,  52.154    ){\circle{2}}
\put(  172.80    ,  55.027    ){\circle{2}}
\put(  175.20    ,  57.900    ){\circle{2}}
\put(  177.60    ,  60.773    ){\circle{2}}
\put(  180.00    ,  63.646    ){\circle{2}}
\put(  182.40    ,  66.519    ){\circle{2}}
\put(  184.80    ,  69.392    ){\circle{2}}
\put(  187.20    ,  72.265    ){\circle{2}}
\put(  189.60    ,  75.138    ){\circle{2}}
\put(  192.00    ,  78.011    ){\circle{2}}
\put(  194.40    ,  80.884    ){\circle{2}}
\put(  196.80    ,  83.757    ){\circle{2}}
\put(  199.20    ,  86.629    ){\circle{2}}
\put(  201.60    ,  89.502    ){\circle{2}}
\put(  204.00    ,  92.375    ){\circle{2}}
\put(  206.40    ,  95.248    ){\circle{2}}
\put(  208.80    ,  98.121    ){\circle{2}}
\put(  211.20    ,  100.99    ){\circle{2}}
\put(  213.60    ,  103.87    ){\circle{2}}
\put(  216.00    ,  106.74    ){\circle{2}}
\put(  218.40    ,  109.61    ){\circle{2}}
\put(  220.80    ,  112.49    ){\circle{2}}
\put(  223.20    ,  115.36    ){\circle{2}}
\put(  225.60    ,  118.23    ){\circle{2}}
\put(  228.00    ,  121.10    ){\circle{2}}
\put(  230.40    ,  123.98    ){\circle{2}}
\put(  232.80    ,  126.85    ){\circle{2}}
\put(  235.20    ,  129.72    ){\circle{2}}
\put(  237.60    ,  132.60    ){\circle{2}}
\put(  240.00    ,  135.47    ){\circle{2}}
\end{picture}}
\end{picture}
\end{center}
\caption{Temperature dependence of the binary and triple reaction rates
given in Table 1 and normalized to the same value at
$\theta=14\times 10^6\ {}^\circ$K.}
\end{figure}
\end{document}